\documentclass[sort&compress,10pt,5p]{elsarticle}
\usepackage{amssymb}
\usepackage{hyperref}

\title{First Large Scale Production of Low Radioactivity Argon From Underground Sources}
\author[princeton]{H. O.~Back\corref{cor1}}
\author[princeton]{F.~Calaprice}
\author[princeton]{C.~Condon}
\author[princeton]{E.~de Haas}
\author[snolab]{R.~Ford}
\author[princeton]{C.~Galbiati}
\author[princeton]{A.~Goretti}
\author[princeton]{T.~Hohmann}
\author[princeton]{An.~Ianni}
\author[princeton]{B.~Loer\fnref{fn1}}
\author[princeton]{D.~Montanari\fnref{fn1}}
\author[princeton]{A.~Nelson}
\author[umass]{A.~Pocar}

\cortext[cor1]{Corresponding author. {email: hback@fnal.gov} \\{Tel.: +1-630-840-2218}, {fax +1-630-840-4263} 
\\{address: Fermilab MS351, P.O. Box 500, Batavia, IL 60510}}
\fntext[fn1]{Now at Fermi National Accelerator Laboratory,\\ P.O.~Box 500, Batavia, IL 60510}

\address[princeton]{Department of Physics, Princeton University, Jadwin Hall, Princeton, NJ 08544}
\address[snolab]{SNOLAB, 1039 Regional Road 24, Creighton Mine \#9, Lively, ON, Canada, P3Y 1N2}
\address[umass]{Department of Physics, 1126 Lederle Graduate Research Tower (LGRT), University of Massachusetts, Amherst, MA 01003}

\renewcommand{\sup}[1]{$^{\mathrm{#1}}$}
\newcommand{\sub}[1]{$_{\mathrm{#1}}$}
\newcommand{\iso}[2]{\sup{#1}$\!${#2}}
\newcommand{\ar}{\iso{39}{Ar}}
\renewcommand{\deg}{\sup{\circ}}
\newcommand{\imgwidth}{\columnwidth}

\begin{document}

\begin{abstract}
We report on the first large-scale production of low radioactivity argon from underground gas wells. Low radioactivity argon is of general interest, in particular for the construction of large scale WIMP dark matter searches and detectors of reactor neutrinos for non-proliferation efforts. Atmospheric argon has an activity of about 1 Bq/kg from the decays of \ar; the concentration of \ar\ in the underground argon we are collecting is at least a factor of 100 lower than this value.

The argon is collected from a stream of gas from a CO$_2$ well in southwestern Colorado with a Vacuum Pressure Swing Adsorption (VPSA) plant. The gas from the well contains argon at a concentration of 400-600 ppm, and the VPSA plant produces an output stream with an argon concentration at the level of 30,000–-50,000 ppm (3–-5\%) in a single pass. This gas is sent for further processing to Fermilab where it is purified by cryogenic distillation. The argon production rate is presently 0.5 kg/day.
\end{abstract}


\maketitle

\section{Introduction}
Argon is a powerful scintillator and an excellent medium for detection of ionization.  Its high discrimination power against minimum ionization tracks, in favor of selection of nuclear recoils, makes it an attractive medium for direct detection of WIMP dark matter~\cite{Benetti2008,Boulay2006,LippincottA}. Argon derived from the atmosphere, however, contains 1 part in 10\sup{15} of the radioactive isotope \ar, which undergoes beta decay (Q=565 keV, t\sub{1/2}=269~y), giving a specific activity of $\sim$1~Bq/kg~\cite{Loosli1983,Benetti2007}.  Both the direct background and pileup from \ar\ decays set limits on the sensitivity and maximum practical size of liquid argon dark matter searches. A source of argon with reduced \ar\ content is necessary to allow sensitive argon-based dark matter searches at the ton-scale and beyond.

The availability of large quantities of argon with low levels of \ar\ may also enable proposed experiments to study neutrinos from high-intensity stopped-pion neutrino sources through neutrino-nucleus elastic scattering with the potential to constrain parameters for non-standard interaction between neutrinos and matter, and to realize precision measurements of the weak mixing angle and of the neutrino magnetic moment~\cite{Scholberg2006}.  Thanks to the excellent separation of nuclear recoils from $\beta/\gamma$ events, low radioactivity argon could also be used for the development of small, portable neutrino detectors to monitor reactor sites for non-proliferation efforts~\cite{Hagmann2004}. Low radioactivity argon could also be used to develop neutron detectors for port security.

Centrifugation and differential thermal diffusion are established methods for \ar /\iso{40}{Ar} isotopic separation, but are impractical for our use because of high cost and slow production rate. Since \ar\ is produced by cosmic ray interactions in the upper atmosphere, principally via the \iso{40}{Ar}(n,2n)\ar\ reaction~\cite{Lehmann1989,Lehmann1991}, gas from underground is a possible source of argon with low levels of \ar. 

As shown in Refs.~\cite{Loosli1989,Lehmann1993}, however, not all underground argon shows a reduced \ar\ activity, due to in-situ production by radiogenic processes driven by $\alpha$-decays in the decay chains of long lived natural uranium and thorium. Indeed, some underground argon has been found to have a higher \ar\ activity than atmospheric argon. Prof. Sujoy Mukhopadhyay at Harvard University suggested that argon gas from the Earth's mantle should have much lower concentrations of \ar, because the concentration of uranium and thorium in the mantle is typically at the ppb level, a thousand times lower than in the crust~\cite{MukhopadhyayA}.  He also noted that the CO\sub2 gas fields in the southwestern part of the U.S., especially the Bravo Dome gas field in New Mexico, had been studied by geologists, and the gas was found to be mainly from the mantle. In 2007, low radioactivity argon was discovered in the National Helium Reserve in Amarillo, TX with an \ar\ concentration a factor of at least 20 below that of atmospheric argon~\cite{Acosta-Kane2008}. Subsequently, gases from the Reliant Dry Ice Plant in Bueyeros, NM and from Kinder Morgan CO2 in Cortez, CO were also found to contain low radioactivity argon. Preliminary measurements of the \ar\ in these gases found concentrations at least 10 times less than in atmospheric argon. Further studies of the underground argon from the Kinder Morgan CO2 have shown that the \ar\ concentration is less than 0.65\% of the \ar\ concentration in atmospheric argon~\cite{Xu2012}.

Table~\ref{tab:argonconcentrations}  shows the amount of argon in the gases from these sites. Details of how the argon from these sites were extracted and tested can be found in Refs.~\cite{Acosta-Kane2008}. Based on the small scale Pressure Swing Adsorption unit used in the initial testing of the gas streams~\cite{Acosta-Kane2008}, a larger Vacuum Pressure Swing Absorbtion plant was built and installed to extract the underground argon gas. Although the large plant was initially run at the Reliant Dry Ice Plant in Bueyeros, NM in 2008, the argon in that gas was exhausted, and the production could not continue. In 2009 the plant was moved to the Kinder Morgan CO2 facility in Cortez, CO, where it has been running since 2010.

\begin{table}
 \caption{Argon concentration in different sources of gas.}
 \centering
 \begin{tabular}{l l}
  \hline \hline
  \textbf{Site} & \textbf{Argon concentration} \\
  \hline 
  National Helium Reserve, & 680 ppm~\cite{Acosta-Kane2008}\\$\quad$Amarillo, TX  \\
  Reliant Dry Ice Plant, & 20-40 ppm~\cite{Cassidy2006,Gilfillan2008,CassidyA}\\$\quad$Bueyeros, NM \\
  Kinder Morgan CO2, & 400-600 ppm\\$\quad$Cortez, CO  \\
  \hline
 \end{tabular}
 \label{tab:argonconcentrations}

\end{table}

\section{Plant design and details}
The large flow of gas available in some wells compensates for the small concentration of argon, so as to make available large masses of argon (10 tons or more) in a reasonable time. However, the low concentration of argon makes the shipment of the gas obtained directly from the well impractical and prohibitively expensive. We have therefore developed a special Vacuum Pressure Swing Adsorption (VPSA) plant to concentrate the argon from underground sources on-site, to a typical argon concentration of a few percent. The use of VPSA technology was inspired by Princeton University's earlier experience operating a Vacuum Swing Adsorption (VSA) radon filter for the clean room used for the construction and assembly of the Borexino nylon vessels~\cite{Pocar2003}.

Pressure (and Vacuum) Swing Adsorption plants work by exploiting the different rates of adsorption of different species of gas at a given partial pressure. “Heavy” gases are characterized by higher adsorption coefficients and are preferentially retained in the adsorbent beds, and thus selectively removed from the input stream. “Light” gases, characterized by small adsorption coefficients, are thus concentrated in the output stream. Once the bed is saturated with “heavy” gases, it can be regenerated by counter-current purging with a portion of the stream of the product gas, or by evacuating the saturated column to high vacuum. By operating two or more columns out of phase, an almost continuous product stream can be maintained. For more details on the VPSA process, see, for instance, Refs.~\cite{Ruthven1994,Knaebel1965,Chiang1996}.

\begin{figure}
 \includegraphics[width=\imgwidth]{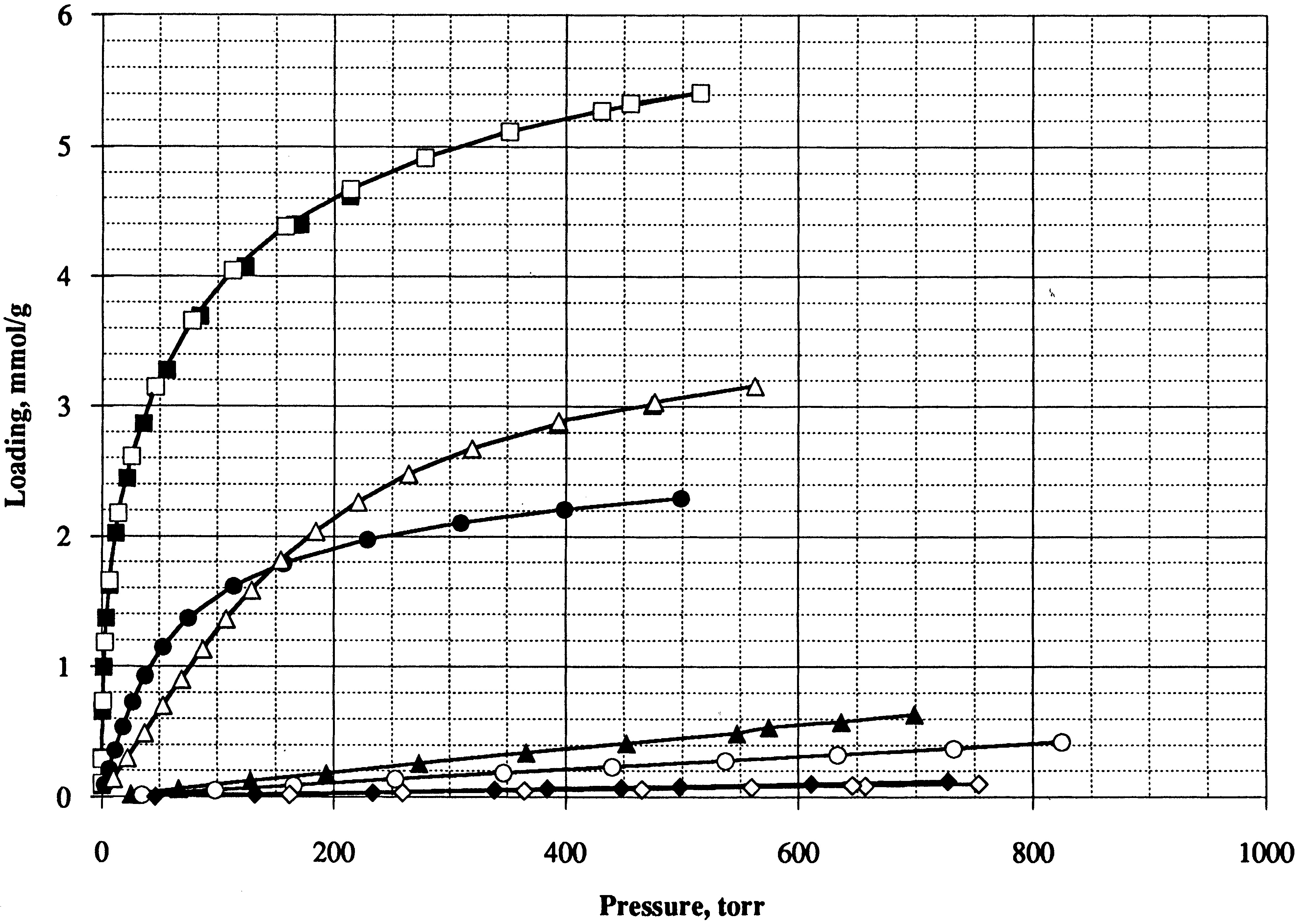}
 \caption{Adsorption isotherms on zeolite NaX: 
 ($\blacksquare$)CO\sub2 at 31.4 \deg C; 
 ($\square$)CO\sub2 at 32.8 \deg C;
 ($\bullet$)SF\sub6 at 31.6 \deg C;
 ($\triangle$)C\sub2H\sub6 at 2.4 \deg C;
 ($\blacktriangle$)CH\sub4 at 31.3 \deg C;
 ($\circ$)N\sub2 at 32.5 \deg C;
 ($\lozenge$)Ar at 31.0 \deg C;
 ($\blacklozenge$)O\sub2 at 33.1\deg C; \cite{Dunne1996}. 
 ``Reprinted with permission from J.A.~Dunne et.~al., Langmuir 12, 5896 (1996). Copyright 1996 American Chemical Society''}
 \label{fig:naxisotherms}
\end{figure}

We have designed and assembled a two-stage VPSA plant optimized to concentrate argon from a portion of the CO\sub2 stream at Kinder Morgan’s Doe Canyon compressor station. As shown in Table~\ref{tab:kindermorgangas}, the gas is dominated by CO\sub2. We selected zeolite NaX as the adsorbent for the first VPSA stage, given its very high selectivity for CO\sub2 over argon~\cite{Dunne1996}. Figure~\ref{fig:naxisotherms}  shows the adsorption isotherms for CO\sub2, CH\sub4, N\sub2, argon, and some other species on zeolite NaX. Given the high selectivity for CO\sub2 relative to argon, we expected the first VPSA unit using NaX zeolites to remove virtually all of the CO\sub2, CH\sub4 and other hydrocarbons, and H\sub2O, boosting the argon concentration by more than an order of magnitude. With the expectation that the gas produced after the first stage would be primarily nitrogen, we installed a second VPSA unit, using zeolite Li-LSX, a lithium-exchanged, low silicate 13X zeolite with optimal selectivity for nitrogen over argon~\cite{Shen2001,Bulow2002,Sebastian2005}.  We remark that helium has a very low adsorption coefficient, lower than argon, for both NaX and Li-LSX zeolites, and thus is concentrated in the product stream along with the argon.

\begin{table}
 \caption{Gas concentrations from the Kinder Morgan Doe Canyon CO\sub2 wells.}
 \label{tab:kindermorgangas}
 \centering
 \begin{tabular}{ll}
  \hline \hline
  \textbf{Gas Type} & \textbf{Well Concentration} \\
  \hline
  Carbon Dioxide & 96\% \\
  Nitrogen & 2.4\% \\
  Methane & 5,700 ppm \\
  Helium & 4,300 ppm \\
  Other hydrocarbons & 2,100 ppm \\
  Water & 1,000 ppm \\
  Argon & 600 ppm \\
  Oxygen & Below sensitivity \\
  \hline
 \end{tabular}

\end{table}

We adopted the following criteria as guidelines for the desired performance of the plant:
\begin{enumerate}
 \item Concentration of argon in the crude stream produced by the plant $>$5\%
 \item Total production rate of argon $>$300~g/day
\end{enumerate}

\begin{figure}
 \includegraphics[width=\imgwidth]{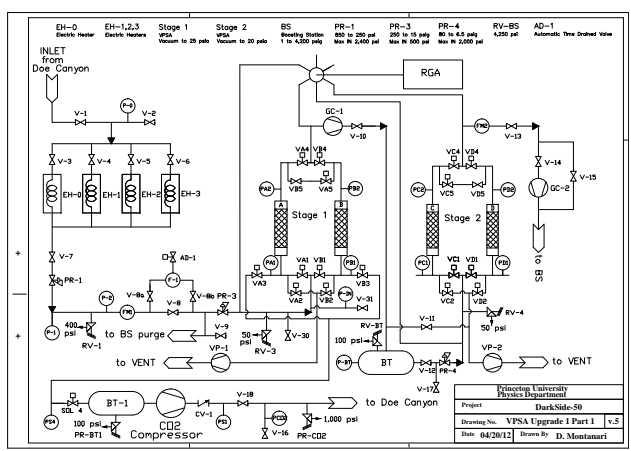}
 \vspace*{-0.12\columnwidth}
 \caption{VPSA plant as-built P\&ID.}
 \label{fig:vpsapid}
\end{figure}

Figure~\ref{fig:vpsapid} is the P\&ID for the VPSA plant as built. The CO\sub2 well head pressure is over 750~psig, whereas our plant operates at a maximum pressure of about 12~psig, which necessitates pressure regulation to step down the pressure to the operation pressure of the plant. The first pressure regulator (PR-1) drops the pressure from the well head pressure to about 240 psig. The temperature of the well gas is the same as the ambient atmosphere, but the temperature drops significantly through the PR-1 due to the Joule-Thompson effect. The pressure drop across PR-1 would be sufficient to cause the CO\sub2 to cool below its liquefaction temperature at 240 psig, and therefore the CO\sub2 can liquefy after the first regulator. To avoid this effect, heaters are used to heat the gas to 70 °C before it passes through PR-1. No heating is required when the gas pressure is further reduced to the VPSA column maximum working pressure through a second pressure regulator (PR-3).

Although the VPSA can remove traces of water, the amount of water in the input gas causes liquid water to accumulate in the gas plumbing due to the cooling through the first regulator. A coalescing gas filter and water trap (F-1) is located immediately after the first pressure regulator. This unit filters any particulate that may exist in the gas stream, and collects liquid water before the adsorption columns. Any residual water vapor in the gas stream is removed by adsorption on the zeolite.

\begin{figure}
 \includegraphics[width=\imgwidth]{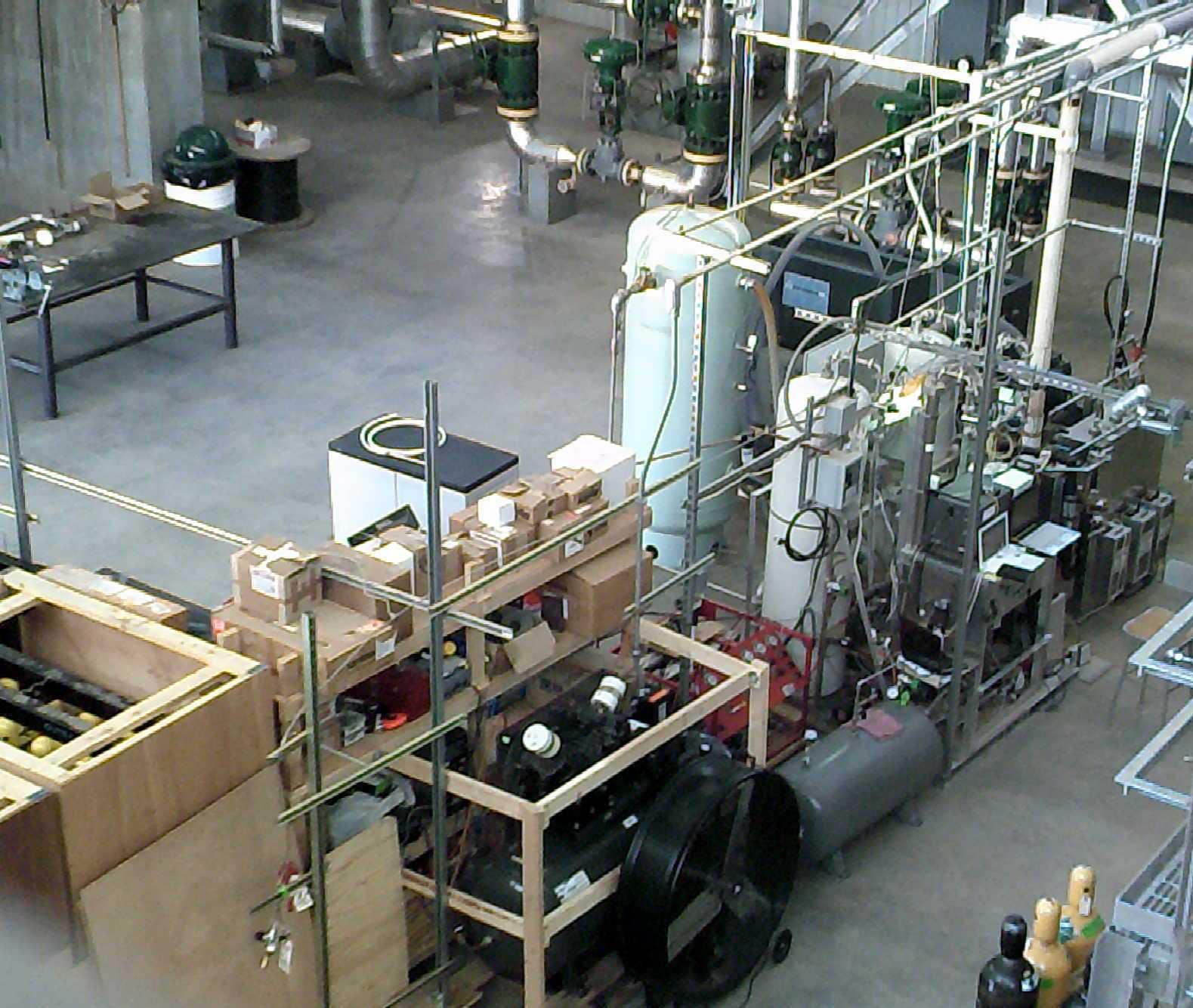}
 \caption{VPSA plant as-build in the Kinder Morgan CO2 facility.}
 \label{fig:plantphoto}
\end{figure}

The VPSA portion of the plant is comprised of two independent VPSA stages operated in series. Each of the two stages runs a routine VPSA process, where each stage consists of two adsorption columns operated 180 degrees out of phase; while one column processes the gas, the other column is being regenerated~\cite{Ruthven1994}. The two VPSA stages each operate between a pressure of $\sim$23 psia and a partial vacuum ($\sim$30–-50 mbar absolute), maximizing recovery of the argon phase. The two stages are separated by a buffer tank (BT), where the product of the first stage is stored at high pressure (40 psig) by use of a gas compressor. The plant was built and assembled on a transportable skid at Princeton University. The columns in stage 1 are 14'' in diameter and 56'' high, and the columns in stage 2 have a 3'' diameter and are 60'' high. We developed and assembled a custom control system, designed to achieve maximum flexibility and to ease the fine tuning of the plant when operating on-site. Figure~\ref{fig:plantphoto} shows a picture of the plant installed at the Kinder Morgan CO2 facility.

The crude argon produced by the VPSA plant is compressed into high pressure cylinders with a 2-stage gas booster that is capable of pressurizing the cylinders to 4000~psig. These gas cylinders can each store the equivalent of 12,000 liters of gas at S.T.P., making storage and shipping of large volumes of gas more economical. They are shipped in manifolded gas transport racks to Fermilab for further processing of the argon using cryogenic distillation~\cite{Back2012A}. 

\begin{figure}
 \includegraphics[width=\imgwidth]{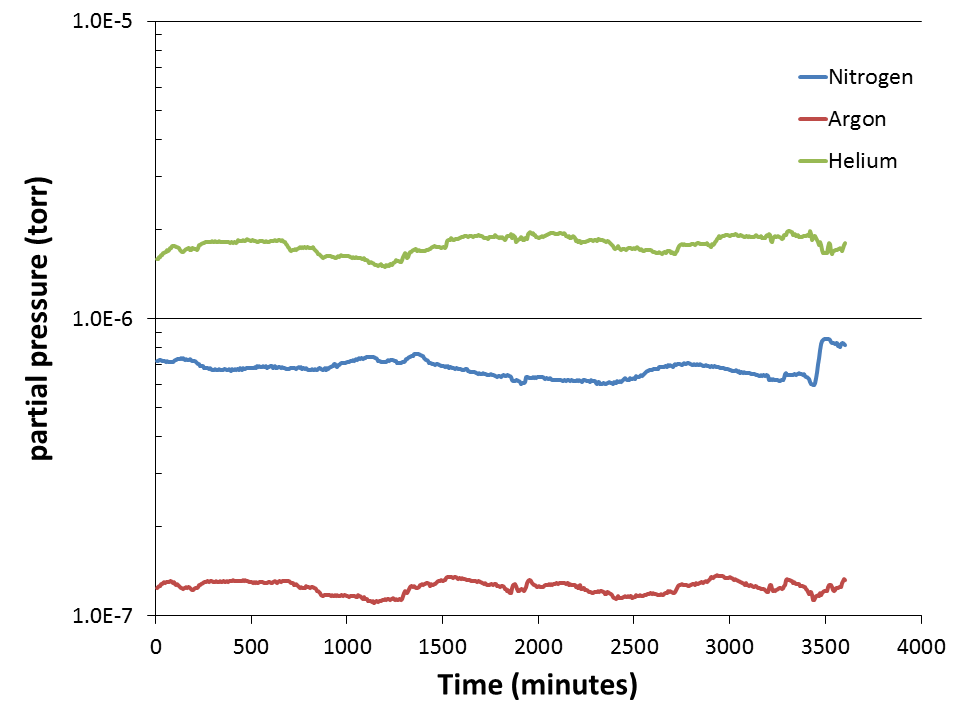}
 \caption{Partial pressure graph of the helium, argon, and nitrogen from the VPSA output for the entire filling cycle of cylinder number 113.}
 \label{fig:exampleugadata}
\end{figure}

At various stages of the gas purification process, the gas can be sampled and analyzed with a dedicated Stanford Research Systems Universal Gas Analyzer (UGA). This UGA serves as the diagnostic tool during plant tuning, and for quality assurance in our final gas product. The sampling points allow measurement of the relative concentrations of the gas constituents in the following locations:
\begin{enumerate}
 \item Inlet of the VPSA plant
 \item Outlet of the first stage
 \item Outlet of the second stage
 \item Gas booster interstage (before filling cylinders)
\end{enumerate}
An example of the UGA data is shown in Figure~\ref{fig:exampleugadata}.

\section{Plant performance}
The VPSA plant was installed at Kinder Morgan's Doe Canyon CO2 Facility in 2009, and was started in early 2010. Commissioning of the plant took about one month, and production began in March 2010. The 2010 year saw the production of approximately 22.6~kg of argon in 45 high pressure cylinders. 

However, in the beginning, the long term stability of the plant and the overall argon production were limited by unforeseen issues and equipment failures. Water in the well gas and in the pneumatic controls of the plant caused valves to fail and affected the adsorption. There were also equipment failures, and the remote location of the facility makes repairs time consuming. With upgrades to the plant, including the addition of water traps, to avoid equipment failures, the production in 2011 was increased. Also, with the plant in a steady running order, the tuning of the plant and its operating procedures were improved. 

Table~\ref{tab:vpsa_gas_concentrations} shows the composition of the VPSA output in 2010 and in 2011. The plant produces low radioactivity argon at a rate of about 0.5~kg/day; however, the average duty factor for the 24 months that the plant has operated is less than 25\%. The duty factor for 2010 was about 12\% primarily due to equipment failure. In 2011 the duty factor was increased to 30\% where the low duty factor was primarily caused by long down times waiting for high pressure cylinder deliveries and some minor equipment failure. As of the end of March 2012 the VPSA plant has produced a total of 85.8 kg of low radioactivity underground argon. 

\begin{table}
 \caption{Composition of the well gas after the VPSA extraction in 2010 and after impovements made in 2011. The fraction of argon is increased significantly from the initial well concentrations listed in Table~\protect{\ref{tab:kindermorgangas}}. Species with concentration listed as ``$\sim$0'' were below the measurement sensitivity of the UGA.}
 \label{tab:vpsa_gas_concentrations}
 \centering
 \begin{tabular}{l c c  }
  \hline
  \hline
  \textbf{Gas Type}  & \multicolumn{2}{p{0.4\columnwidth}}{\textbf{Concentration after VPSA extraction}} \\
  & \textbf{2010} & \textbf{2011} \\
  \hline
  Carbon Dioxide & $\sim$0 & $\sim$0 \\
  Nitrogen & 70\% & 40\% \\
  Methane & $\sim$0 & $\sim$0 \\
  Helium & 27.5\% & 55\% \\
  Other hydrocarbons  & $\sim$0 & $\sim$0 \\
  Water & $\sim$0 & $\sim$0 \\
  Argon & 2.5\% & 5\% \\
  Oxygen & $\sim$0 & $\sim$0 \\
  \hline
\end{tabular}
\end{table}

\section{Conclusions}
We had established with previous sampling campaigns that some underground gas wells contain argon with low levels of \ar\ with respect to the atmospheric concentration~\cite{Acosta-Kane2008}. With this present work, we have demonstrated the practicality of extracting underground argon from gas streams containing minute concentrations of argon (a few hundreds of parts per million) on the scale required for the development of large-scale dark matter detectors. We have developed a technique suitable for processing large flow rates of gas, capable of concentrating the argon phase by more than a factor of 10 in a single pass. This technique has been demonstrated on a primarily CO\sub2 stream, but it is applicable to a variety of naturally occurring gas streams.

We have shown that it is possible to concentrate traces of argon from an underground stream into a crude argon product stream with argon concentrations up to 5\%. We have demonstrated that this plant is already capable of producing $\sim$0.5~kg per day of low radioactivity argon, with a duty factor better than 30\%. Operation of the plant is cost-effective, and the plant will be used to produce a few hundred kilograms of low radioactivity argon for WIMP dark matter searches.

\section{Acknowledgements}

This work was supported in part by National Science Foundation grants PHY NSF-0704220, PHY NSF-0811186, and PHY NSF-1004072, and by Canda Foundation for Innovation grant CFI \#20299. The construction and operation of the two-stage VPSA purification plant was made possible by seed funds granted by the Physics Department of Princeton University and by the University Research Board of Princeton University.

Support for Henning Back at Princeton University was provided by Mark Boulay and Art McDonald of Queen's University, Kingston, Canada, through grants from the Canada Foundation for Innovation and the Ministry of Research and Innovation of the Province of Ontario.

We thank C.\ Callan, L.\ Page, D.\ Marlow, and A.S.\ Smith for their support and for useful discussions.

We are very grateful to Kinder Morgan CO2 for hosting us at their Doe Canyon Facility in Cortez, CO, and special thanks to all of the employees at Kinder Morgan CO2 for making this work possible.


\begin{thebibliography}{99}
 \bibitem{Benetti2008}
 P.~Benetti et.~al. (WARP Collaboration), Astropart.~Phys.~\textbf{28}, 495 (2008).
 \bibitem{Boulay2006}
 M.G.~Boulay and A.~Hime, Astropart.~Phys.~\textbf{25}, 179 (2006).
 \bibitem{LippincottA}
 W.H.~Lippincott et.~al., pre-print arXiv:0801.1531.
 \bibitem{Loosli1983}
 H.H.~Loosli, Earth Plan.~Sci.~Lett.~\textbf{63}, 51 (1983).
 \bibitem{Benetti2007}
 P.~Benetti et.~al. (WARP Collaboration), Nucl.~Instr.~Meth.~A \textbf{574}, 83 (2007).
 \bibitem{Scholberg2006}
 K.~Scholberg, Phys.~Rev.~D \textbf{73}, 033005 (2006).
 \bibitem{Hagmann2004}
 Hagmann and A.~Bernstein, IEEE Trans.~Nucl.~Sci.~\textbf{51}, 2151 (2004).
 \bibitem{Lehmann1989}
 B.E.~Lehmann and H.H.~Loosli, in Proceedings of the 6th International Symposium on Water-Rock Interaction, 3-6 August 1989, Malvern, UK, edited by D.L.~Miles (A.A.~Balkema, 1989), pp.~429432.
 \bibitem{Lehmann1991}
 B.E.~Lehmann and H.H.~Loosli, in Applied Isotope Hydrogeology: A Case Study in Northern Switzerland, edited by F.J. Pearson et.~al. (Elsevier, 1991), pp.~239 296.
 \bibitem{Loosli1989}
 H.H.~Loosli, B.E.~Lehmann, and W.~Balderer, Geochim.~Cosmochim.~Acta \textbf{53}, 1825 (1989)
 \bibitem{Lehmann1993}
 B.E.~Lehmann, S.N.~Davis, and J.T.~Fabryka-Martin, Water Resources Researcg \textbf{29}, 2027 (1993)
 \bibitem{MukhopadhyayA}
 S.~Mukhopadhyay (private communication)
 \bibitem{Acosta-Kane2008}
 Acosta-Kane et.~al. Nucl.~Instr.~Meth.~A \textbf{587}, 46 (2008).
 \bibitem{Xu2012}
 J.~Xu, et.~al. pre-print arXiv:1204.6011.
 \bibitem{Cassidy2006}
 M.~Cassidy. Occurrence and Origin of Free Carbon Dioxide Gas Deposits in the Earths Continental Crust. Ph.D. Thesis, University of Houston, Texas (2006).
 \bibitem{Gilfillan2008}
 S.M.V.~Gilfillan et.~al., Geochim.~Cosmochim.~Acta \textbf{72}, 1174 (2008).
 \bibitem{CassidyA}
 M.~Cassidy, Private Communication.
 \bibitem{Pocar2003}
 A.~Pocar, Low Background Techniques and Experimental Challenges for Borexino and its Nylon Vessels. PhD. Thesis, Princeton University (2003).
 \bibitem{Ruthven1994}
 M.~Ruthven, S.~Farooq, and K.S.~Knaebel, \textit{Pressure Swing Adsorption}, CVS Publishers, Inc., 1994.
 \bibitem{Back2012A} 
 H.O.~Back et.~al., pre-print arXiv:1204.6061.
 \bibitem{Knaebel1965}
 K.S.~Knaebel, F.B.~Hill, Chem.~Eng.~Sci.~\textbf{40}, 2351 (1965).
 \bibitem{Chiang1996}
 A.S.T.~Chiang, Chem.~Eng.~Sci.~\textbf{51}, 207 (1996).
 \bibitem{Dunne1996}
 J.A.~Dunne et.~al., Langmuir \textbf{12}, 5896 (1996).
 \bibitem{Shen2001}
 D.~Shen et.~al., Microporous and Mesoporous Materials \textbf{48}, 211 (2001).
 \bibitem{Bulow2002}
 M.~B\"{u}low, D.~Shen, and S.~Jale, Applied Surface Science \textbf{196}, 157 (2002).
 \bibitem{Sebastian2005}
 J.~Sebastian, S.A.~Peter, R.V.~Jasra, Langmuir \textbf{21}, 11220 (2005).
\end{thebibliography}
\end{document}